\documentclass[3p,times,twocolumn]{elsarticle}

\usepackage{ecrc}

\volume{00}

\firstpage{1}

\journalname{Nuclear and Particle Physics Proceedings}

\runauth{M. Koles\'{a}r, J. \v{R}\'{i}ha}

\jid{nppp}

\jnltitlelogo{Nuclear and Particle Physics Proceedings}

\usepackage{amssymb}

\usepackage[figuresright]{rotating}

\newcommand{\no}{\noindent}
\newcommand{\vsp}[1]{\vspace{#1}}
\newcommand{\hsp}[1]{\hspace{#1}}

\newcommand{\myeq}[3]{\vspace{#2} \begin{equation} \hspace{#1} #3 \end{equation} \vspace{0cm}}
\newcommand{\myeqn}[3]{\vspace{#2} \begin{displaymath} \hspace{#1} #3 \end{displaymath} \vspace{0cm}}

\begin{document}

\begin{frontmatter}



\dochead{}

\title{Application of Bayesian statistics to \\$\eta$-meson decay constant in $\chi$PT}


\author[label1]{Mari\'{a}n Koles\'{a}r\corref{cor1}}
\address[label1]{Institute of Particle and Nuclear Physics, Faculty of Mathematics and Physics, Charles University, Prague, Czech republic}
\cortext[cor1]{Speaker}
\author[label1]{Jaroslav \v{R}\'{i}ha}

\begin{abstract}
The decay constant of the $\eta$-meson in the framework of 'resummed' chiral
perturbation theory is discussed. A theoretical prediction is
compared to the available determinations. Compatibility of these determinations with the latest 
fits of the $SU(3)$ low energy coupling constants is investigated.  Preliminary results for the obtained constraints on the low energy coupling constants $L_5^r$ and $L_4^r$, using Bayesian statistical approach, are presented.
\end{abstract}




\end{frontmatter}



\section{Decay constants of the light pseudoscalar mesons}

Decay constants of the light pseudocalar meson nonet can be introduced in terms of the QCD axial currents 

\myeq{1.5cm}{0cm}{\label{axial currents}
		i p_{\mu} F_P^a = \langle\,0\,|\,A_{\mu}^a(0)\,|\,P\,\rangle,}

\no where	$A_{\mu}^a=	\bar{q}\gamma_\mu \gamma_5 \lambda^a q$. The pion and kaon decay constant take a straightforward form in the isospin limit and are very well know from either experimental data or lattice simulations \cite{PDG2018, Aoki:2019cca}. Quite a lot of work has been devoted to the $\eta$-$\eta'$ sector, where substantial mixing occurs, see e.g. \cite{Leutwyler:1997yr,Feldmann:1998vh,Benayoun:1999au,Escribano:2005qq,Klopot:2012hd,Guo:2015xva,Escribano:2015yup,Bali:2017qce,EMTC:2019xmh}. The results span a range of values, some of which are not compatible with others. 
		
In the framework of $SU(3)_L\times SU(3)_R$ chiral perturbation theory ($\chi$PT) \cite{Gasser:1984gg}, the chiral expansion of the pseudoscalar meson octet in the isospin limit can be written in the following way
								
\myeq{-0.75cm}{0cm}{\label{Fpi}
		F_{\pi}^2 = F_0^2( 1-4\mu_{\pi}-2\mu_K) + 8 m_\pi^2 (L_4^r(r+2)+L_5^r) + 
										\Delta_{F_{\pi}}}
\myeqn{-0.75cm}{-0.5cm}{
		F_K^2 = F_0^2(1-\frac{3}{2}\mu_{\pi}-3\mu_K-\frac{3}{2}\mu_{\eta})}
\myeq{1cm}{-0.5cm}{\label{FK}
						+ 8 m_\pi^2 (L_4^r(r+2)+\frac{1}{2}L_5^r(r+1)) + \Delta_{F_K}}
\myeq{-0.75cm}{-0.5cm}{\label{Feta}
		F_{\eta}^2 = F_0^2(1-6\mu_K) + 8 m_\pi^2 (L_4^r(r+2)+\frac{1}{3}L_5^r(2r+1)) + 
								   \Delta_{F_{\eta}}.}	
									
\no This form is obtained directly from the generation functional of two point Green functions in the logic of 'resummed' approach to $\chi$PT \cite{DescotesGenon:2003cg}. $\Delta_{F_P}$ denote the sum of all higher orders, the so-called higher order remainders. $\mu_P=m_P^2/(32\pi^2F_0^2) \ln(m_P^2/\mu^2)$ are the chiral logarithms, where $m_P$ are the pseudoscalar masses at leading order. In particular, $m_\pi^2=2B_0\hat{m}$. $r=m_s/\hat{m}$ is the ratio of strange and light quark masses.

The $SU(3)$ decay constant $F_\eta$ is defined identically to $F_\eta^8$ in (\ref{axial currents}) and can thus be related to the mixing parameters in the $U(3)$ octet-singlet basis						
									
\myeq{1.5cm}{0cm}{F_\eta = F_{\eta}^8 = F_8\cos\vartheta_8.}

\no For the purpose of this work, we will use two values of $F_\eta^8$ as our input
		
\myeq{0.5cm}{0cm}{F_\eta^8 = (1.18 \pm 0.02)F_\pi \quad \mathrm{(EGMS15) \label{EGMS15}}}	
\myeq{0.5cm}{-0.5cm}{F_\eta^8 = (1.38 \pm 0.05)F_\pi \quad \mathrm{(EF05) \label{EF05}}}	

\no EGMS15 \cite{Escribano:2015yup} is a recent determination which is representative of lower values of this observable. On the other hand, EF05 \cite{Escribano:2005qq} lies on the opposite end of the spectrum and is an example of a very high value of $F_\eta^8$. Reported uncertainties are quite low in both cases and thus these results are essentially incompatible with each other. 

As can be seen from (\ref{Fpi}-\ref{Feta}), chiral expansions of the decay constants up to next-to-leading order depend only on the two leading order and two NLO low energy constants (LECs) - $F_0$, $B_0$ and $L_4^r$, $L_5^r$, respectively. The most recent standard $\chi$PT fit \cite{Bijnens:2014lea} provides two different sets for the NLO LECs (at $\mu=M_\rho$)

\myeq{-0.5cm}{0cm}{10^3L_4^r \equiv 0.3,\hsp{1.15cm} 10^3L_5^r = 1.01\pm 0.06 \quad \mathrm{(BE14)\ }}
\myeq{-0.5cm}{-0.5cm}{\label{FF14}
		10^3L_4^r = 0.76\pm 0.18,\ 10^3L_5^r = 0.50\pm 0.07 \mathrm{\quad (FF14)\ }}

\no The main fit (BE14) fixes $L_4^r$ by hand, in order to ensure the expected suppression in the large $N_c$ limit. FF14 (free fit) releases this constraint. As can be seen, the obtained values are quite different. The difference is much less pronounced for the LO LECs:

\myeq{0cm}{0cm}{\label{BE14_LO}
		F_0 = 71\mathrm{MeV},\ Y = m_\pi^2/M_\pi^2 = 1.055 \quad \mathrm{(BE14)\ }}
\myeq{0cm}{-0.5cm}{\label{FF14_LO}
		F_0 = 64\mathrm{MeV},\ Y = m_\pi^2/M_\pi^2 = 0.937 \quad \mathrm{(FF14)\ }}

The purpose of this work is twofold - first, we will show that the 'resummed' $\chi$PT framework leads to a simple, but robust prediction for $F_\eta$. Then we will use the two values of $F_\eta^8$ (\ref{EGMS15}-\ref{EF05}) as an input and use a Bayesian statistical approach to obtain constraints on the higher order remainders and the NLO LECs. We will compare these results with the two versions of the fit \cite{Bijnens:2014lea} (BE14 and FF14) and thus check the compatibility of the various values of $F_\eta$ and NLO LECs.

\section{Bayesian statistical analysis}	

We use a statistical approach based on the Bayes' theorem \cite{DescotesGenon:2003cg,Kolesar:2017xrl}																							
\myeq{0.25cm}{0cm}{
		P(X_i|\mathrm{data}) = 
		\frac{P(\mathrm{data}|X_i)P(X_i)}{\int \mathrm{d}X_i\,P(\mathrm{data}|X_i)P(X_i)},}
		
\no where $P(X_i|\mathrm{data})$ is the probability density of the theoretical parameters, denoted as $X_i$, having a specific value given the experimental input. 

In the case of independent experiments, $P(\mathrm{data}|X_i)$ is the known probability density of obtaining the observed values of the observables $O_k$ in a set of experiments with uncertainties $\sigma_k$ under the assumption that the true values of $X_i$ are known		
																		
\myeq{-0.5cm}{0cm}{P(\mathrm{data}|X_i) = \prod_k\frac{1}{\sigma_k\sqrt{2\pi}}\,
		\mathrm{exp}\left[-\frac{(O_k^\mathrm{exp}-O^\mathrm{th}_k(X_i))^2}{2\sigma_k^2}\right].}
		
$P(X_i)$ are prior probability distributions of $X_i$. We use them to implement the theoretical assumptions and uncertainties connected with our parameters.

\section{Assumptions}		

For the LO LECs $F_0$ and $B_0$, we use the same theoretical constraints as in \cite{Kolesar:2017xrl}, which define our priors for these parameters. Their approximate range then is  
																				
\myeq{1cm}{0cm}{0 < Y < Y_{\mathrm{max}} \simeq 2.5}
\myeq{1cm}{-0.5cm}{0 < Z < Z(2) = 0.86 \pm 0.01,}	

\no where $Y=2B_0\hat{m}$, $Z=F_0^2/F_\pi^2$ and $Z(2)=F(2)^2/F_\pi^2$. $F(2)$ is the $SU(2)$ pion decay constant in the chiral limit.

As for the NLO LECs $L_4^r$ and $L_5^r$, we limit them to the range (at $\mu=770$MeV)

\myeq{1.5cm}{0cm}{L_4^r, L_5^r \in (0,2\times 10^{-3}).}

We estimate the higher order remainders statistically, based on general arguments about the convergence 
of the chiral series \cite{DescotesGenon:2003cg}

\myeq{1cm}{0cm}{\delta_{F_P} = \Delta_{F_P}/F_P^2 = 0.0\pm 0.1. \label{delta}}

\no We implement this by normal distributions. The remainders are thus limited only statistically, not by any upper bound.

We use the lattice QCD average \cite{Aoki:2019cca} for the value of the strange-to-light quark mass ratio $r$ 

\myeq{2cm}{0cm}{r= 27.43 \pm 0.31.}

Finally, the inputs for the pion and kaon decay constants are \cite{PDG2018}

\myeq{-0.5cm}{0cm}{F_\pi = 92.21 \pm 0.15\ \mathrm{MeV},\ F_K = 110 \pm 0.28\ \mathrm{MeV}.}

\section{Results}

We will employ several ways of dealing with the system of equations (\ref{Fpi}-\ref{Feta}). As a first step, it is possible to eliminate $F_0$, $L_4^r$ and $L_5^r$ by simple algebraic manipulations and thus obtaining a single equation

\myeqn{-0.65cm}{0cm}{
		 F_{\eta}^2 = \frac{1}{3}\Big[4F_K^2-F_{\pi}^2 + \frac{M_{\pi}^2 Y}{16\pi^2}
                      (\ln \frac{m_{\pi}^2}{m_K^2} + (2r+1)\ln \frac{m_{\eta}^2}{m_K^2})}
\myeq{2.25cm}{-0.25cm}{     
		+ 3F_\eta^2\delta_{F_{\eta}} - 4F_K^2\delta_{F_K} + 
                      F_\pi^2\delta_{F_{\pi}}\Big].}
																
\begin{figure}[t]
\begin{center}																						
\includegraphics[scale=0.6]{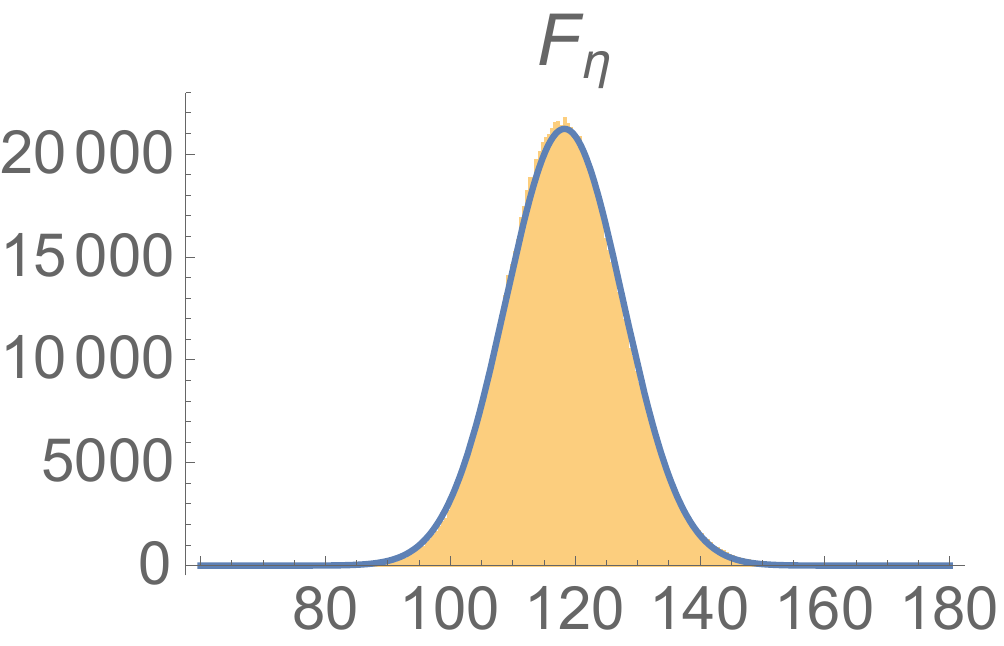}																						
\end{center}
\caption{Prediction for $F_\eta$ ($10^6$ points). Gaussian fit overlaid.}
\label{f1}
\end{figure}

\no The equation depends, beyond the remainders $\delta_{F_P}$, only on a single parameter $Y$ and the dependence is very small, as already noted in \cite{DescotesGenon:2003cg} and \cite{Kolesar:2008fu}. A histogram of $10^6$ numerically generated theoretical predictions, depending on the assumptions listed in the previous section, is depicted in Fig.\ref{f1}. A Gaussian fit leads to a value 

\myeq{0cm}{0cm}{F_\eta = 118.3\pm 9.4\ \mathrm{MeV} = (1.28\pm0.10)F_\pi.}

\no This is an improved prediction over \cite{Kolesar:2008fu} and lies in between the values of EGMS15 (\ref{EGMS15}) and EF05 (\ref{EF05}), discussed above. 

\begin{figure}[b!]
\begin{center}																						
\includegraphics[scale=0.5]{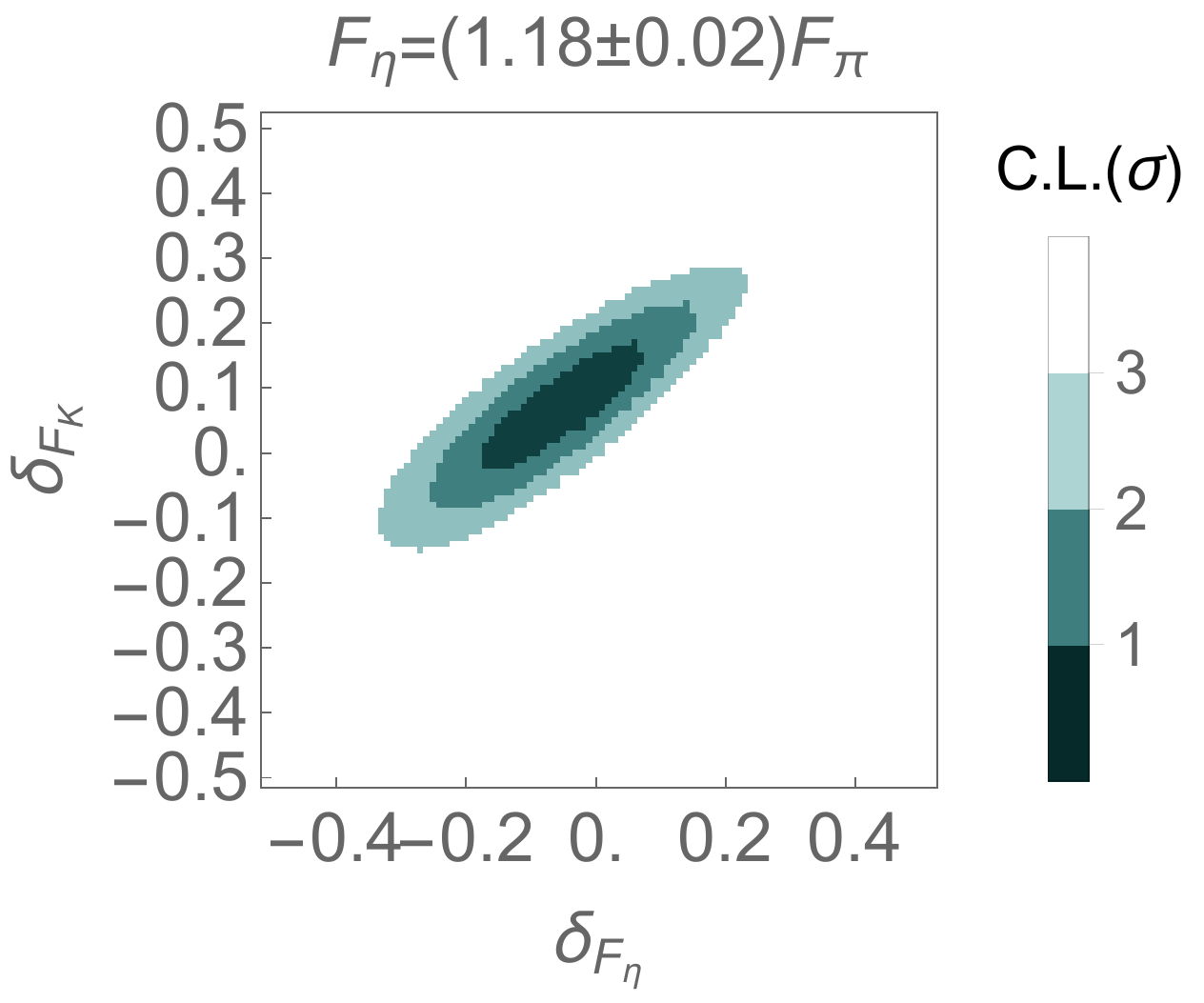}																						
\includegraphics[scale=0.5]{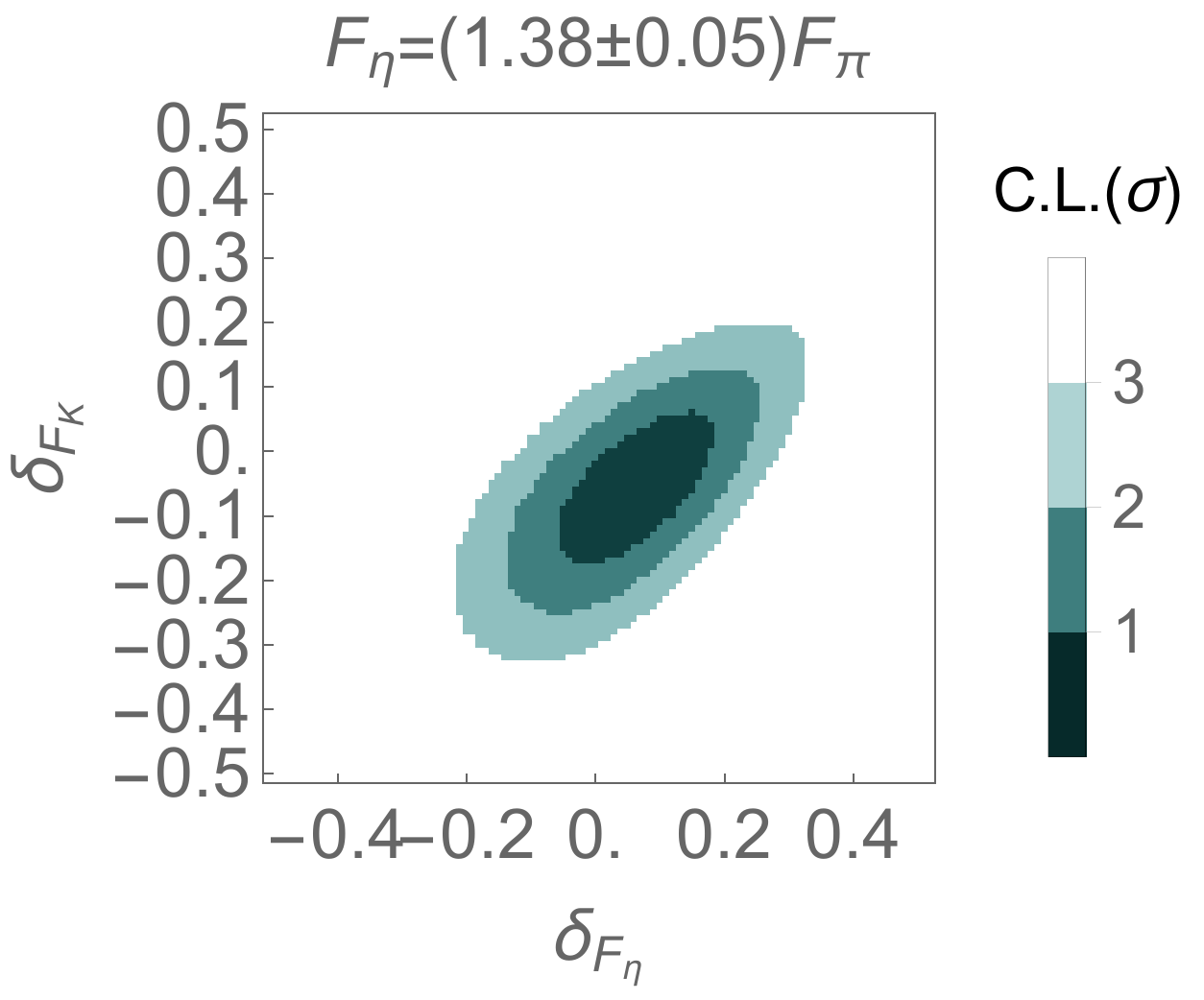}																						
\end{center}
\caption{Constraints on higher order remainders from $F_\eta$.}
\label{f2}
\end{figure}	

Next, we can use EGMS15 and EF05 as inputs and employ the Bayesian statistical approach to extract information about the remainders. A contour plot with confidence levels can be found in Fig.\ref{f2}, which leads to

\myeq{-0.75cm}{0cm}{\delta_{F_K} = 0.07 \pm 0.06,\ \delta_{F_\eta} = -0.06 \pm 0.08\ \ \mathrm{(EGMS15)}}
\myeq{-0.75cm}{-0.5cm}{\delta_{F_K} = -0.06 \pm 0.08,\ \delta_{F_\eta} = 0.05 \pm 0.08\ \ \mathrm{(EF05).}}

\no  We can see that the values are compatible with the prior assumption (\ref{delta}). We can also compare this result with the NNLO contributions for $F_K$ obtained in \cite{Bijnens:2014lea}

\myeq{0.5cm}{0cm}{F_K/F_\pi = 1 + 0.176 + 0.023\quad \mathrm{(BE14)}}
\myeq{0.5cm}{-0.5cm}{F_K/F_\pi = 1 + 0.121 + 0.077\quad \mathrm{(FF14).}}

\no As can be seen, both are positive, while EF05 implies a negative remainder $\delta_{F_K}$. It should be noted, however, that the work \cite{Bijnens:2014lea} uses a different form of the chiral expansion and thus this can only be taken as an indication that lower values of $F_\eta$ might be better compatible with the fits BE14/FF14.

\begin{figure}[t]
\begin{center}																						
\includegraphics[scale=0.6]{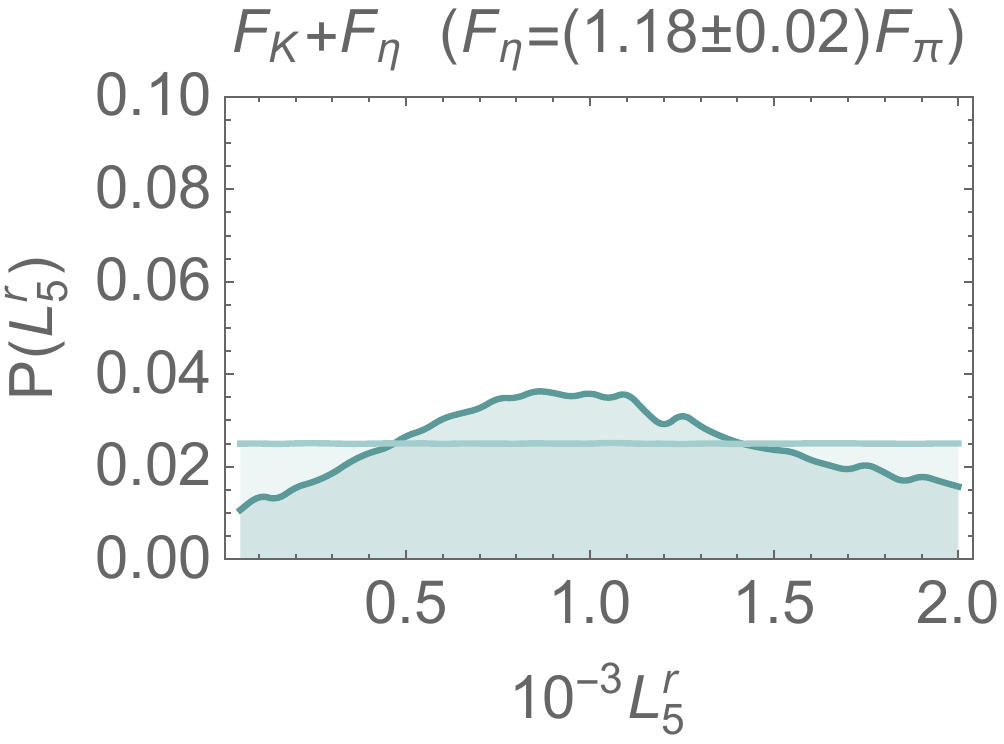}
\includegraphics[scale=0.6]{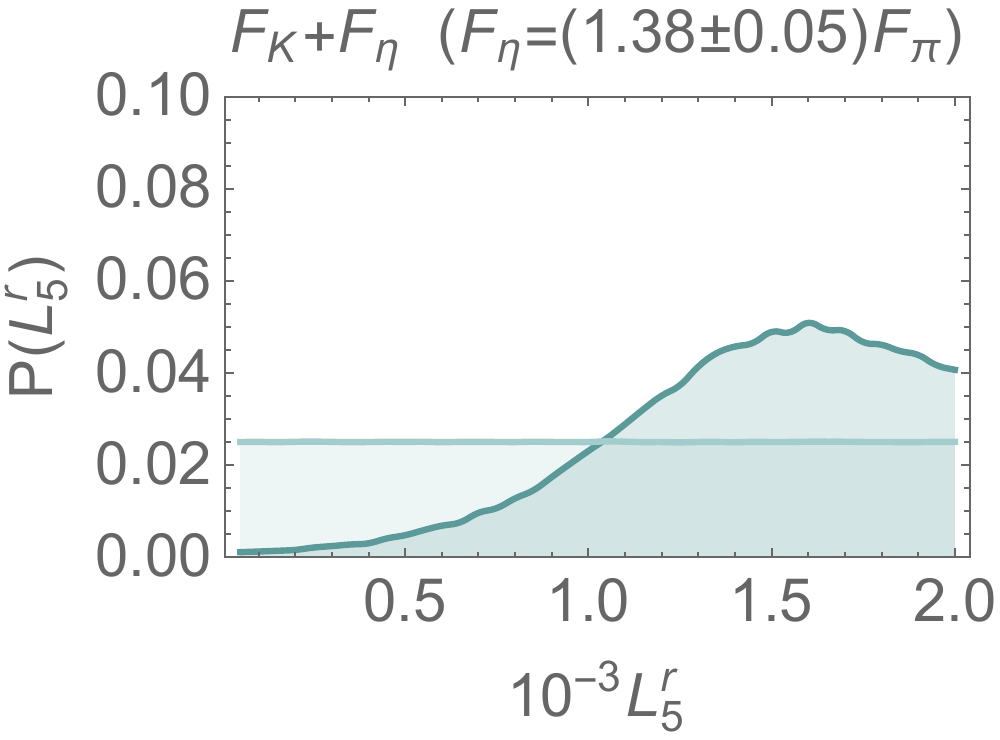}																							
\end{center}
\caption{PDF for $L_5^r$ from $F_K$ and $F_\eta$ (transparent- prior).}
\label{f3}
\end{figure}

As a second possibility, we can algebraically eliminate $F_0$ and $L_4^r$ from equation (\ref{Fpi}), which leads to a system of two equations for $F_K$ and $F_\eta$, now depending on $Y$, $L_5^r$ and the remainders $\delta_{F_P}$. The obtained constraints on $L_5^r$, from $2\cdot10^7$ numerically generated predictions, are shown in Fig.\ref{f3}.

Lower values of $L_5^r$ are preferred in the case of EGMS15 ($F_\eta$=$(1.18 \pm 0.02)F_\pi$), but the result is not statistically significant. However, in the case a high value of $F_\eta$ (EF05), we obtain a limit

\myeq{0.5cm}{0cm}{\label{L5_EF05}
L_5^r>0.7\cdot10^{-3}\ \mathrm{at\ 2\sigma\ C.L.\quad (EF05)},}

\no which is incompatible with the value from the fit FF14 (\ref{FF14}), which is $L_5^r=(0.5\pm0.07)\cdot10^{-3}$.

\begin{figure}[t]
\begin{center}																						
\includegraphics[scale=0.6]{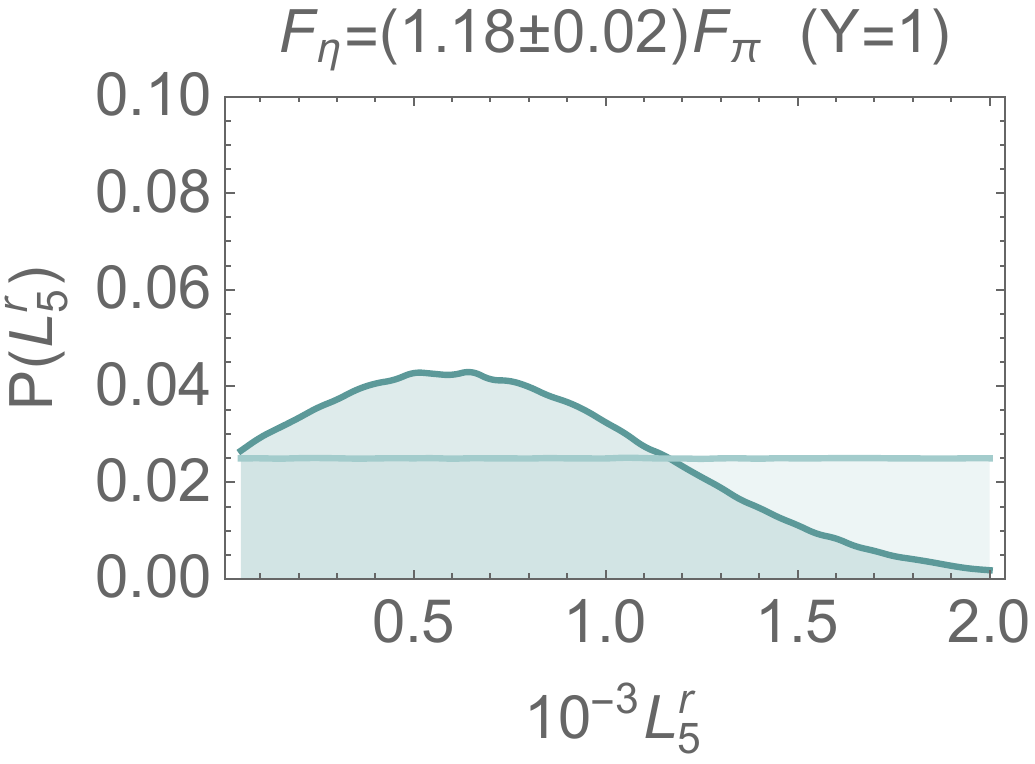}																						
\end{center}
\caption{PDF for $L_5^r$ from $F_\eta$ at $Y=1$ (transparent- prior).}
\label{f4}
\end{figure}	

In the next step, we will add an additional assumption. There is little difference in the value of $Y=m_\pi^2/M_\pi^2$ for the two fits BE14 (\ref{BE14_LO}) and FF14 (\ref{FF14_LO}). Hence we will take a look at the consequences of assuming $Y=1$.

Fig.\ref{f4} shows the updated probability density function for $L_5^r$ for EGMS15. We obtain an upper bound

\myeq{-0.25cm}{0cm}{L_5^r<1.6\cdot10^{-3}\ \mathrm{at\ 2\sigma\ C.L.}\ \ (Y=1)\ \ \mathrm{(EGMS15)}.}

\no The lower bound for EF05 (\ref{L5_EF05}) stills holds for $Y=1$.

\begin{figure}[t]
\begin{center}																						
\includegraphics[scale=0.6]{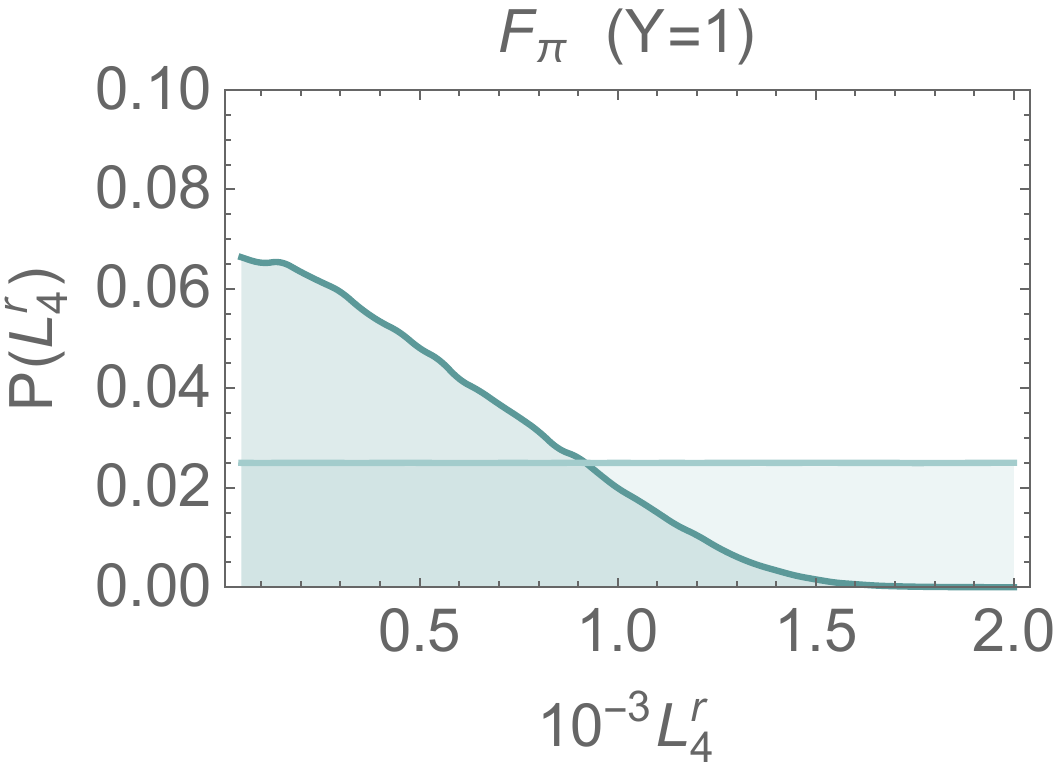}																						
\end{center}
\caption{PDF for $L_4^r$ from $F_\pi$ at $Y=1$ (transparent- prior).}
\label{f5}
\end{figure}

As the last option, we will try to extract information on $L_4^r$. In this case, we use equation (\ref{FK}) to eliminate $L_5^r$. Interestingly, the strongest constraint in this case is obtained from the chiral expansion for $F_\pi$ (\ref{Fpi}), shown in Fig.\ref{f5}. We get

\myeq{0.5cm}{0cm}{L_4^r<1.2\cdot10^{-3}\ \mathrm{at\ 2\sigma\ C.L.}\ \ (Y=1).}

\vsp{-0.5cm}
\section{Summary}

We have applied the Bayesian statistical approach to the sector of light pseudoscalar mesons in the framework of 'resummed' $\chi$PT, while using two different inputs for the $\eta$-meson decay constants. We have investigated the compatibility of these inputs with the most recent fits of the relevant $\chi$PT low energy constants. Our results can be shortly summarized as

\begin{itemize}								
	\item[$\bullet$] Our prediction is $F_\eta = (1.28\pm0.10)F_\pi$.\vsp{-0.25cm}
	\item[$\bullet$] $F_\eta$=$(1.38\pm0.05)F_\pi$ (EF05) implies negative higher order 
												corrections to $F_K$, possibly in contradiction with the fits BE14/FE14.\vsp{-0.25cm}
	\item[$\bullet$] $F_\eta=(1.38\pm0.05)F_\pi$ (EF05) implies $L_5^r>0.7\cdot10^{-3}$ at 2$
												\sigma$ C.L., incompatible with the fit FE14.\vsp{-0.25cm}
	\item[$\bullet$] Y=1 and $F_\eta=(1.18\pm0.02)F_\pi$ (EGMS15) implies $L_5^r<1.6
												\cdot10^{-3}$ at 2$\sigma$ C.L.\vsp{-0.25cm}
	\item[$\bullet$] Y=1 implies $L_4^r<1.2\cdot10^{-3}$ at 2$\sigma$ C.L. (from $F_\pi$).				
\end{itemize}

\no {\bf Acknowledgment:} The work was financially supported by
The Czech Science Foundation (project GACR no.18-17224S).




\nocite{*}
\bibliographystyle{elsarticle-num}
\bibliography{Bibliography}







\end{document}